\documentclass{webofc}
\usepackage[varg]{txfonts}   
\usepackage{hyperref}
\usepackage{url}
\hypersetup{colorlinks=true,citecolor=blue,urlcolor=blue,linkcolor=blue}

\begin{document}
\title{Investigating subnucleonic structures via new measurements of incoherent J/$\psi$ photoproduction in ultra-peripheral Pb--Pb collisions with ALICE}
\author{\firstname{Vendulka} \lastname{Humlová}\inst{1}\fnsep\thanks{\email{vendulka.filova@cern.ch}} on behalf of the ALICE Collaboration}
\institute{Faculty of Nuclear Sciences and Physical Engineering, Czech Technical University in Prague, Czech Republic}

\abstract{
Ultra-peripheral collisions of heavy ions provide a unique environment to study the gluon structure of nuclei through photon-induced reactions. In particular, the incoherent photoproduction of J/$\psi$ vector meson is sensitive to event-by-event fluctuations of the gluon field at nucleon and subnucleonic scales. We report new ALICE measurement of incoherent J/$\psi$ production in Pb--Pb collisions at $\sqrt{s_{\mathrm{NN}}}=5.02$~TeV, differential in both photon--nucleus energy and Mandelstam-$|t|$. At low $|t|$, the cross section rises with increasing photon--nucleus energy, while at larger $|t|$ the growth is suppressed with respect to the trend at low $|t|$. This measurement represents the first study of incoherent photoproduction at the LHC performed as a function of both $W_{\gamma \mathrm{Pb},n}$ and $|t|$.
}
\maketitle

\section{Introduction}
\label{intro}
Deep-inelastic scattering experiments have shown that, at small Bjorken-$x$, the partonic structure of hadrons is dominated by gluons, whose density rises with energy approximately as a power law~\cite{HERA}. Quantum chromodynamics (QCD) predicts that such growth is eventually balanced by gluon recombination---a phenomenon known as saturation. Finding clear experimental signatures of saturation is one of the major challenges of high-energy nuclear physics, and a central motivation for future facilities such as the Electron--Ion Collider~\cite{EIC}.

Diffractive photoproduction of vector mesons provides a sensitive probe of gluon densities~\cite{Ryskin,KleinReview}. In this process, a quasi-real photon emitted by a nucleus fluctuates into a $c\bar c$ dipole which scatters off the target via gluon exchange, producing a vector meson. When the photon interacts with the entire nucleus, the process is referred to as coherent. If instead the photon couples to a single nucleon inside the nucleus, the process is incoherent. The incoherent channel can be further divided into elastic, where the struck nucleon remains intact, and dissociative, where the nucleon breaks up into a higher-mass system. In incoherent interactions the nucleus as a whole breaks up, usually leading to the emission of forward neutrons in the same direction as the incoming target. Coherent production probes the average nuclear density, while incoherent production is sensitive to event-by-event fluctuations at nucleon and subnucleonic levels.

The kinematics of heavy vector meson (J/$\psi$) production, where the hard scale is set by the J/$\psi$ mass, can be reconstructed from the measured rapidity ($y$) and transverse momentum ($p_{\mathrm{T}}$) of the J/$\psi$. The squared four-momentum transfer is very well approximated as $|t| \approx p_{\mathrm{T}}^2$ in the kinematic region of this analysis, while the photon--nucleus centre-of-mass energy is given by $W_{\gamma \mathrm{Pb},n}^2 = M_{J/\psi}\,\sqrt{s_{\mathrm{NN}}}\,e^{\pm y},$
where $M_{J/\psi}$ is the J/$\psi$ mass and $\sqrt{s_{\mathrm{NN}}}$ the nucleon--nucleon centre-of-mass energy.
These kinematic relations also connect to the geometric interpretation of the process: Mandelstam-$|t|$ is related through a Fourier transform to the impact parameter, making it possible to study the transverse spatial distribution of gluons at different size scales. 

ALICE recently reported the first multi-differential measurement of incoherent J/$\psi \rightarrow \mu^{+}\mu^{-}$ photoproduction as a function of both photon--nucleus energy and $|t|$, and observed, at high energy, a suppression of the cross section growth at large $|t|$ compared to low $|t|$~\cite{ALICEincoherent}.

\section{Experimental setup and analysis}
\label{sec-setup}
The analysis uses Pb--Pb UPC data collected in 2018 at $\sqrt{s_{\mathrm{NN}}}=5.02$~TeV, corresponding to an integrated luminosity of about 533~$\mu$b$^{-1}$. The ALICE apparatus is described in detail in Ref.~\cite{ALICEexp}.

The ALICE Muon Spectrometer covers the pseudorapidity region $-4 < \eta < -2.5$ and is designed to reconstruct J/$\psi \to \mu^+\mu^-$ decays. It consists of a hadron absorber, five tracking stations inside a dipole magnet, and two trigger stations. This system ensures efficient muon identification and precise momentum measurement down to low transverse momentum.

The Zero Degree Calorimeters (ZDCs) detect forward neutrons emitted in the beam direction. By classifying events according to neutron detection (0nXn or Xn0n), the ZDCs determine the photon-emitter direction and separate interactions at low ($ W_{\gamma \mathrm{Pb},n} \sim 20,30$~GeV) and high ($ W_{\gamma \mathrm{Pb},n} \sim 633$~GeV) photon energies. Neutrons emitted in incoherent photoproduction signal the photon direction, identifying which nucleus emitted the photon and which served as the target. V0A and AD are scintillator detectors that provide vetoes against hadronic activity. The V0A detector is located on the side opposite to the Muon Spectrometer, while the AD detectors are placed on both sides of the interaction point, ensuring the selection of the diffractive process.

The V0A detector, located on the side opposite to the Muon Spectrometer, is used as the primary veto 
against hadronic activity. The AD counters, installed on both sides of the interaction point, serve as 
additional veto detectors.

The J/$\psi$ signal is extracted from unbinned extended likelihood fits to the dimuon invariant mass distributions, using a double-sided Crystal Ball function for the signal and an exponential for the background. Contaminating contributions from coherent J/$\psi$ and feed-down from $\psi(2S)$ are determined from a fit to the transverse momentum distribution of J/$\psi$ yield. The fit model consists of Monte Carlo STARlight~\cite{Starlight} templates and the H1 parametrization of the $p_{\mathrm{T}}$ spectrum for nucleon-dissociation diffractive events~\cite{H1}.

Cross sections are measured in three $|t|$ intervals ($0.09<|t|<0.36$, $0.36<|t|<0.81$, and $0.81<|t|<1.44$~GeV$^2$) and multiple photon--nucleus energies. The dominant systematic uncertainties arise from muon trigger efficiency, the selection on the number of fired V0C cells, and signal extraction~\cite{ALICEincoherent}.

\section{Results and discussion}
\label{sec-results}
\begin{figure}[!t]
\centering
\includegraphics[width=0.7\textwidth]{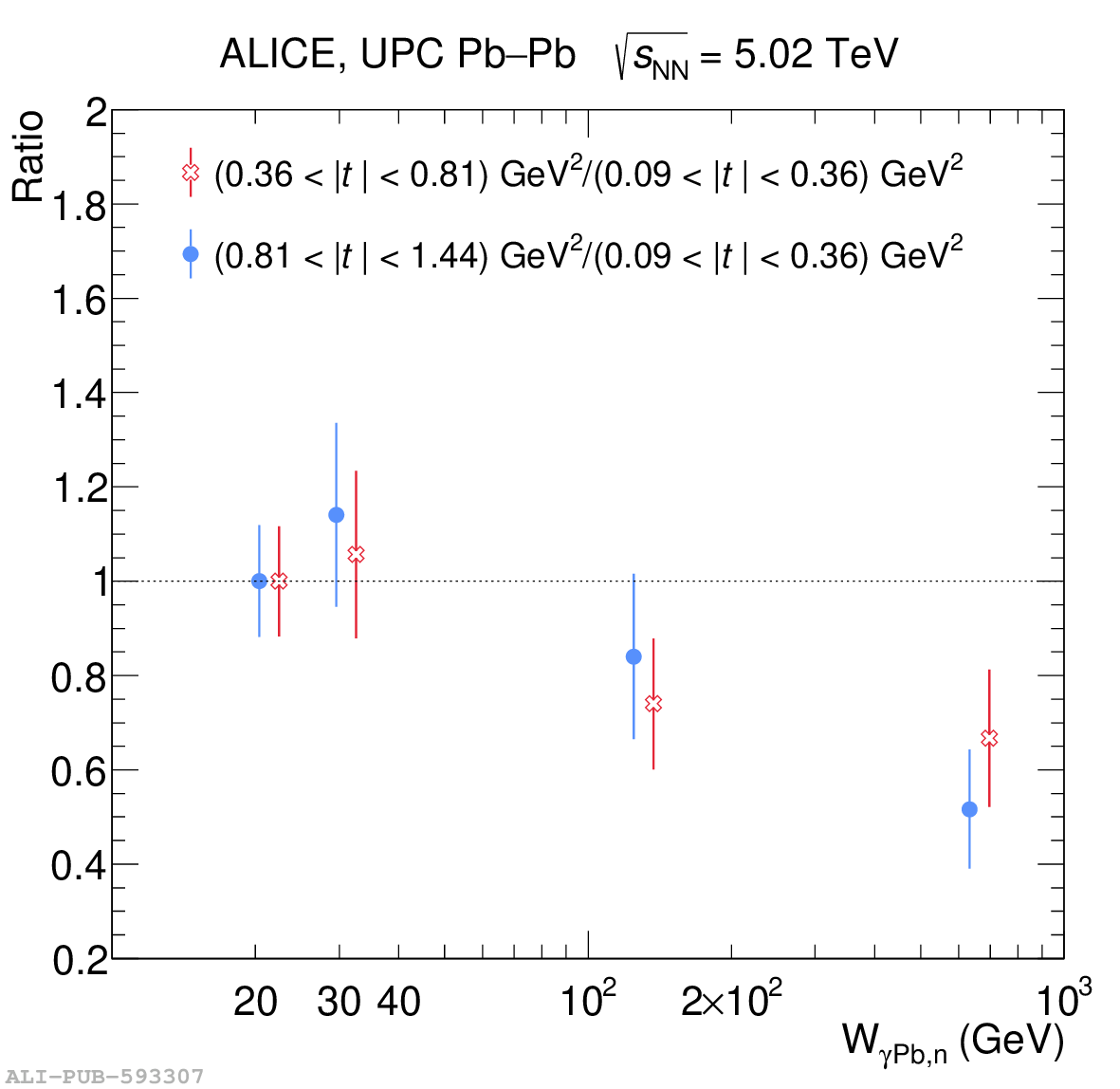}
\caption{Ratios of the cross sections at different Mandelstam-$t$ ranges normalized such that the lower energy ratio is one. The vertical line denotes the statistical and $|t|$-uncorrelated systematic uncertainties added in quadrature.}
\vspace{-13pt} 
\label{fig:ratio}
\end{figure}
The incoherent J/$\psi$ photoproduction cross section rises as the energy increases. At high $|t|$, probing subnucleonic structures, the growth of the cross section is suppressed with respect to the lower $|t|$ values. 

The suppression is quantified via ratios of cross sections at large and small $|t|$ regions, normalized at low $W_{\gamma \mathrm{Pb},n}$. The ratio is shown in Fig.~\ref{fig:ratio}. At $W_{\gamma \mathrm{Pb},n}=633$~GeV, the ratio between the highest and lowest $|t|$ bins is $0.52 \pm 0.13$, deviating from unity by more than three standard deviations~\cite{ALICEincoherent}. This demonstrates that the cross section growth is significantly reduced when probing smaller spatial gluonic configurations.

\begin{figure}[!t]
\centering
\includegraphics[width=0.98\textwidth]{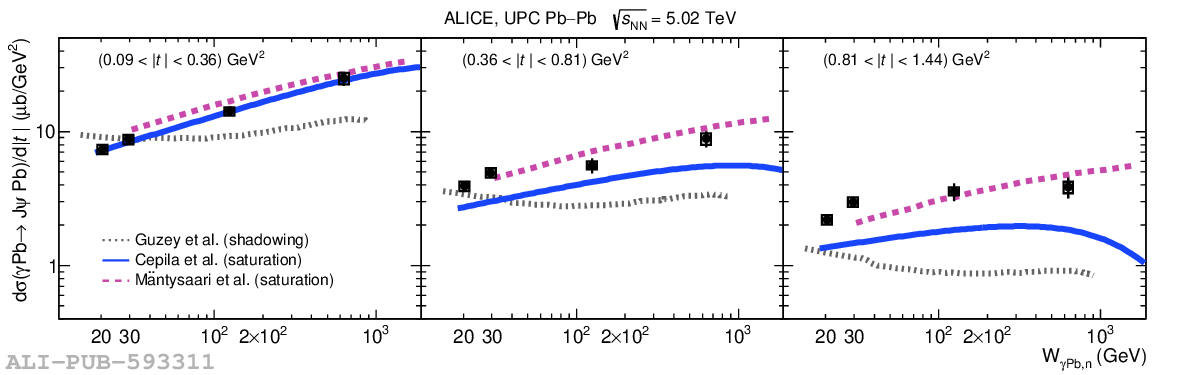}
\caption{The energy dependence of the cross section of incoherent J/$\psi$ photonuclear production at three ranges of the Mandelstam-$t$ compared with model predictions. The shadowing-based model of Guzey et al.~\cite{Guzey} is shown with a dot-dashed line. Two saturation-based predictions are also shown: from M{\"a}ntysaari et al.~\cite{Mantysaari} with a dashed line and from Cepila et al.~\cite{Cepila} with a solid line.}
\vspace{-13pt} 
\label{fig:models}
\end{figure}

In order to explore the origin of this suppression, data are compared with model predictions, as shown in Fig.~\ref{fig:models}. The shadowing-based model by Guzey et al.~\cite{Guzey}, reproduces coherent J/$\psi$ production at high energy~\cite{ALICEcohEnergy} but fails to describe the incoherent data. Saturation-based models incorporating subnucleonic fluctuations provide a better description. 

In the energy-dependent hot-spot approach of Cepila et al.~\cite{Cepila}, the reduction arises from the overlap of an increasing number of gluonic hot spots at high energies, which lowers the variance of fluctuations. The approach of Mäntysaari et al.~\cite{Mantysaari} instead predicts a decreasing power-law exponent with increasing $|t|$, leading to a slower growth of the cross section. Both of the considered saturation-based approaches reproduce the qualitative features of the data, with the model by Mäntysaari et al. providing a better description.
\section{Conclusion and outlook}
\label{sec-conclusion}
The first multi-differential measurement of incoherent J/$\psi$ photoproduction in Pb--Pb UPCs as a function of both energy and Mandelstam-$|t|$ has been presented. The energy dependence of this suppression of cross section growth at large $|t|$, compared with the models discussed, suggests that gluon saturation plays a significant role as smaller spatial scales are probed. 

With the increased statistics expected in LHC Runs 3 and 4, and thanks to the extended coverage provided by the new forward detectors such as FoCal, ALICE will extend these studies to even smaller Bjorken-$x$ values. Incoherent production will remain a powerful tool for studying fluctuations in the nuclear gluon field.


\end{document}